\documentclass[pre,floatfix,notitlepage,nofootinbib]{revtex4-1}

\usepackage[utf8]{inputenc}
\usepackage{amsmath}
\usepackage{amssymb}
\usepackage{lipsum}
\usepackage{bm}
\usepackage{graphicx}
\usepackage{graphics}
\usepackage[dvipdf]{epsfig}
\usepackage{subcaption}
\usepackage{float}
\usepackage{wrapfig}
\usepackage{url}
\usepackage{epstopdf} 
\usepackage{xcolor}
\usepackage{tensor} 
\usepackage{setspace}

\newcommand{\bEB}{{\mathbf{E}_{\text{B}}}}
\newcommand{\bEBt}{{\mathbf{E}_{\text{Biot}}}}
\newcommand{\bEBl}{{\mathbf{E}_{\text{Bell}}}}
\newcommand{\bB}{{\mathbf{B}}}
\newcommand{\bC}{{\mathbf{C}}}
\newcommand{\bU}{{\mathbf{U}}}
\newcommand{\bV}{{\mathbf{V}}}

\begin{document}
\title{Stretch formulations and the Poynting effect in nonlinear elasticity}

\author{\;Eduardo Vitral\;}
\email{Email: evitral@unr.edu}
\affiliation{Department of Mechanical Engineering, University of Nevada,
    1664  N.  Virginia  St.  (0312),  Reno,  NV  89557-0312,  U.S.A.}

\begin{abstract}

  The second invariant of the left Cauchy-Green deformation tensor $\bB$ (or right $\mathbf{C}$) has been argued to play a fundamental role in nonlinear elasticity. Generalized neo-Hookean materials, which depend only on the first invariant, lead to universal relations that conflict with experimental data, fail to display important mechanical behaviors (such as the Poynting effect in simple shear), and may not provide a satisfactory link with the mesoscale. However, the second invariant term is not a higher order strain contribution to the energy, which lead us to reflect on what is incomplete about neo-Hookean materials. Instead of the usual Cauchy-Green elastic formulation, we investigate this matter from the perspective of left stretch $\bV=  \sqrt{\bB}$ and Bell strain $\bEBl = \bV-\mathbf{I}$ formulations. Invariants of these tensors offer a different interpretation than those of $\bB$ and are linked to different classes of materials. The main example we adopt is a general isotropic energy quadratic in Bell strains, the quadratic-Biot material. Despite being quadratic in stretch like neo-Hookean, this material presents both the classic and reverse Poynting effect in simple shear, whose direction switches as a function of the constant conjugate to the second invariant of $\bEBl$. Its second normal stress also presents a local maximum as a function of the amount of shear, a transition that is not observed in a Mooney-Rivlin solid. Moreover, even the Varga model, linear in Bell strains, presents Poynting in simple shear, which poses the question of why this is not true for a model linear in Green-Lagrange strains. Pure torsion of a solid cylinder is also discussed, particularly how the behavior of the resultant axial force contrasts between the different formulations. 
  
\end{abstract}
\date{\today}


\maketitle


\section{Introduction}

Simple hyperelastic materials present a strain energy density function $\mathcal{W}$ that, for isothermal deformations, depends only on the current deformation gradient $\mathbf{F}$, which completely determines the stress tensor. For isotropic elastic solids, the material symmetry group of the reference undistorted configuration is the special orthogonal group $\textrm{SO}(3)$, and thus, by the polar decomposition of $\mathbf{F}$ and objectivity, the stress becomes a function of the left stretch $\bV$ only, or, equivalently, of the left Cauchy-Green deformation tensor $\bB = \bV^2$~\cite{truesdell2004non}. Therefore, the most general form of the energy $\mathcal{W}$ for these solids can be written in terms of the three invariants of the stretch or Cauchy-Green deformation tensors. When adopting a dependency on invariants $i_i^\bB$ of $\bB$, one popular class of materials is the generalized incompressible neo-Hookean class, for which the energy is a function of only the first invariant $\mathcal{W}_1 = \mathcal{W}(i_1^\bB)$, a simple functional form that also appears from the kinetic theory of rubber. Recent discussions on micro to macro connections in these materials, shortcomings of the $\mathcal{W}_1$ class, and possible generalizations can be found in Puglisi and Saccomandi~\cite{puglisi2016multi}, Destrade et. al~\cite{destrade2017methodical}, and Anssari-Benam and Bucchi~\cite{anssari2021generalised}.

Despite the popularity of $\mathcal{W}_1$ models, a number of studies emphasize the importance of the second invariant $i_2^{\bB}$ in the modeling of nonlinear elastic materials -- see for example Wineman~\cite{wineman2005some}, Horgan and Smayda~\cite{horgan2012importance}, and Anssari-Benam et al.~\cite{anssari2021central}. A powerful tool for evaluating the nonlinear character and how appropriately a material class corresponds to experimental evidence is the derivation of universal relations~\cite{beatty1987class,horgan1999simple,anssari2021central,murphy2022inverted}. Hence, one strong argument for incorporating $i_2^\bB$ in Cauchy-Green type energies is the fact that experimental data conflicts with universal relations based on $\mathcal{W}_1$~\cite{horgan1999simple,destrade2015extreme}. Moreover, the second invariant is argued as essential to provide a better fit for experimental data, and to model mechanical behaviors that generalized neo-Hookean materials fail to capture, such as the Poynting effect in isochoric simple shear under plane stress (i.e. the normal stress perpendicular to the shearing direction required to maintain this deformation). While the addition of $i_2^\bB$ solves the conflict with universal relations and recovers desirable mechanical behaviors, the underlying issues of why ``incomplete'' generalized neo-Hookean materials fail in the first place to model important responses, and what is special about $i_i^\bB$ from a mathematical modeling standpoint, have not been explored in detail.

The present paper proposes to look into this problem from a different perspective by adopting a stretch formulation~\cite{ogden1972large,rivlin2004note} for the strain energy and derived stresses, instead of the usual formulation based on the Cauchy-Green tensor. More precisely, we will employ a measure of deformation linear in stretch, known as the Bell strain $\bEBl = \bV-\mathbf{I}$ (spatial counterpart of the referential Biot strain), and write the energy in terms of its invariants $i_i^\bEB$~\cite{vitral2022quadratic}. While both formulations have their own advantages and can be translated from one form to the other~\cite{rivlin2004note}, the Cauchy-Green one is historically preferred. This happens because equations formulated in terms of $\bB$ and its invariants often possess a simpler form, and their associated strains, the Green-Lagrange and Euler-Almansi tensors, can be easily written as a function of metrics and bases of the problem. On the other hand, stretch formulations often lead to more convoluted equations, and to complications such as tensor square roots, although explicit equations for the stretch tensors in terms of their own invariants are available~\cite{hoger1984determination,ting1985determination,vitral2022quadratic}. However, the role played by the invariants $i_i^\bEB$ differs significantly when compared to the one played by $i_i^\bB$ in their respective formulations, which provides new insights into canonical nonlinear elastic problems such as simple shear of a cuboid and pure torsion of a cylinder.

When it comes to stretch type energies, we focus our attention on the isotropic quadratic-Biot material~\cite{vitral2022quadratic,lurie1968theory}, which, similarly to the classic neo-Hookean material, is quadratic in stretch. However, the former is a two constant general quadratic energy in Bell strains that can be constructed from a systematic expansion in eigenvalues of $\bEBl$. Therefore, particularly for small finite strains, we are interested in contrasting the mechanical response predicted by these two different energies. The quadratic-Biot material has been recently adopted to derive reduced plate and shell energies~\cite{vitral2022dilation,vitral2022energies}, which avoids the undesirable mixing between stretching and bending contents introduced when the reduction is performed for certain energies quartic in stretches~\cite{irschik2009continuum,oshri2017strain,wood2019contrasting}, such as Saint Venant-Kirchhoff. It also leads to a complete two constant bending energy for an isotropic material, instead of the one constant bending energy derived from neo-Hookean.

Another question is how a subclass of $\mathcal{W}_1$ governed by $(i_1^\bB-3)^n$ fares against a stretch class of energies of the type $(i_1^\bEB)^n$, where $n$ is an integer. Among the stretch class, we have the Varga model for $n=1$, and a one constant quadratic-Biot model for $n=2$. In other words, this is a comparison between models constructed with powers of a strain quadratic in stretch, with an energy limited to even powers of stretch~\cite{bufler1995drilling,vitral2022quadratic}, and those built with powers of a strain linear in stretch. We emphasize that a particular energy is independent of the formulation: it can always be rewritten in terms of another set of invariants (although translating invariants of $\bV$ into those of $\bB$ is a convoluted task, involving a quartic equation~\cite{hoger1984determination}). The present work is concerned with contrasting \textit{different} simple functional forms (linear, quadratic) constructed from different sets of invariants or strain measures. As discussed by Hoger~\cite{hoger1999second}, a constitutive theory that is of a certain order in a strain measure will not be of the same order in a different measure, leading to distinct mechanical behaviors. That work provides a detailed derivation of second order theories in Biot strain, whereas here we provide a comparison between theories that are linear or quadratic in Bell strain and those with strain measures based on invariants of $\bB$.

We introduce the quadratic-Biot energy in Section~\ref{sec:biot}, adopting the incompressibility constraint, and write the Bell and Cauchy stress tensors in terms of the invariants of the Bell strain. In Section~\ref{sec:shear} we reformulate the problem of simple shear with traction free lateral condition on the basis of invariants of the left stretch and Bell strain, and discuss the consequences of adscititious inequalities to the constants of the quadratic-Biot material. While for this homogeneous deformation the Poynting effect is absent in generalized neo-Hookean materials, the quadratic-Biot material not only displays classic (positive) Poynting, but also the reverse (negative) effect inside the allowable range of the material's parameters. This effect is present for stretch type energies even when there is no functional dependence of $i_2^\bEB$. Additionally, it is shown that a quadratic-Biot material presents shear hardening and a transition in the second normal stress as a function of the amount of shear, which is not the case for a Mooney-Rivlin material. This section closes with a reflection on what is recovered on Cauchy-Green formulations when $i_2^\bB$ is added from the point of view of $\bEBl$, a primitive strain linear in stretch. The second example is pure torsion of a solid cylinder, in Section~\ref{sec:torsion}, which we again reformulate on the Bell strain basis. We remark that no evident relationship between the resultant applied moment and axial force can be found for a class of stretch type energies, as is the case for $\mathcal{W}_1$. From the resultant axial force, we also compare the Poynting effect between the Varga, neo-Hookean, and quadratic-Biot materials as a function of the angle of twist.

\section{Isotropic quadratic-Biot material}
\label{sec:biot}

Consider a reference unstressed body where the position vector $\mathbf{X}$ locates material points. When the body undergoes a deformation, the reference position $\mathbf{X}$ is mapped into $\mathbf{x}$, describing the deformed configuration. The deformation gradient $\mathbf{F}$ satisfies $d\mathbf{x} = \mathbf{F}\cdot d\mathbf{X}$, so that $\mathbf{F} = \textrm{Grad}\,\mathbf{x}$. By the polar decomposition, $\mathbf{F}$ can be uniquely decomposed as
\begin{equation}
  \mathbf{F} = \bV\cdot\mathbf{Q} = \mathbf{Q}\cdot\bU\,,
\end{equation}
where $\mathbf{Q}\in \textrm{SO}(3)$, and the stretch tensors $\bV$ and $\bU$ are symmetric positive-definite. The left (spatial) and right (referential) Cauchy-Green deformation tensors are $\bB = \mathbf{F}\cdot\mathbf{F}^\top$ and $\bC = \mathbf{F}^\top\cdot\mathbf{F}$, respectively, and are related to the stretch tensors by $\bB = \bV^2$ and $\bC = \bU^2$. Note that the eigenvalues of $\bV$ and $\bU$ coincide, which are the principal stretches $\lambda_i$, with $i\in\{1,2,3\}$, whereas those of $\bB$ and $\bC$ are $\lambda_i^2$. The Green-Lagrange $\frac{1}{2}(\bC-\mathbf{I})$ and Euler-Almansi $\frac{1}{2}(\mathbf{I}-\bB^{-1})$ tensors are often adopted as measures of strain, which are quadratic in stretch.

While the elastic energy function for isotropic materials is typically written in terms of invariants of the left $i_i^\bB$ (or right $i_i^\bC$) Cauchy-Green tensor, we can also cast it as a function of invariants of stretches~\cite{steigmann2002invariants,rivlin2004note} or of the symmetric Bell $\bEBl = \bV - \mathbf{I}$ (or Biot $\bEBt = \bU-\mathbf{I}$) strain~\cite{beatty1992deformations}. The latter is a primitive measure of strain, linear in stretch, and can be adopted as a small expansion parameter to derive a general quadratic-stretch elastic energy~\cite{vitral2022quadratic}, which presents both even and odd powers of stretches. In contrast, hyperelastic models based on $i_i^\bB$, such as Saint Venant-Kirchhoff and Mooney-Rivlin materials (both quartic in stretch), are limited to even powers of stretches~\cite{bufler1995drilling}.  Among these models, we find the generalized neo-Hookean class of materials, which is independent of $i_2^\bB$, and thus incomplete; not only they are deficient in describing general mechanical responses, but even for small deformations they do not correspond to any systematic expansion in strains.\footnote{While this text focus on the spatial $\bV$ and $\bB$ tensors, arguments are equivalent for formulations based on the referential $\bU$ and $\bC$ tensors due to parity of eigenvalues and invariants.}

A general isotropic quadratic energy function of the Bell strain is of the form
\begin{equation}
    \mathcal{W}(i_1^\bEB,i_2^\bEB) = c_1\,(i_1^\bEB)^2 + c_2\,i_2^\bEB \,,
    \label{eq:wbiot}
\end{equation}
where $c_1$ and $c_2$ are constant material parameters and the principal invariants of $\bEBl$ are
\begin{eqnarray}
  \nonumber
  i_1^\bEB &=& \textrm{Tr}\,\bEBl = \Delta_1+\Delta_2+\Delta_3\,,
  \\[2mm]
  i_2^\bEB &=& \frac{1}{2}\left[\textrm{Tr}^2\,\bEBl - \textrm{Tr}\,(\bEBl^2)\right] = \Delta_1\Delta_2+\Delta_2\Delta_3+\Delta_1\Delta_3\,,
               \label{eq:invariants}
  \\[2mm]
  \nonumber
  i_3^\bEB &=& \textrm{Det}\,\bEBl = \Delta_1\Delta_2\Delta_3\,.
\end{eqnarray}
The eigenvalues of $\bEBl$ have a clear physical interpretation~\cite{hoger1999second}: they are principal strains, the distance of principal stretches from unity, $\Delta_i = \lambda_i-1$. If one enforces $c_1 \ge -c_2/3$ and $c_2 \le 0$, then~\eqref{eq:wbiot} is a convex function, with a positive-definite Hessian in terms of $\Delta_i$. However, a stronger constraint on these constants can be imposed due to restrictions on response function, which will be discussed in Section~\ref{sec:rest}.

A material governed by~\eqref{eq:wbiot} has been labeled as ``semilinear'' by Lurie~\cite{lurie1968theory}, and in two-dimensions as ``harmonic'' by John~\cite{john1960plane,steigmann1988stability}. Due to the lack of an universally adopted name for~\eqref{eq:wbiot}, we refer to it as a \textit{quadratic-Biot} material. The energy~\eqref{eq:wbiot} can also be written in terms of invariants of the left stretch $\mathbf{V}$,
\begin{equation}
  \overline{\mathcal{W}}(i_1^{\mathbf{V}},i_2^{\mathbf{V}}) = c_1\,(i_1^{\mathbf{V}}-3)^2 + c_2\,(i_2^{\mathbf{V}}-3)-2c_2\,(i_1^{\mathbf{V}}-3) \,,
  \label{eq:biotv}
\end{equation}
where $i_i^\bV$ are analogously defined as~\eqref{eq:invariants} in terms of $\bV$ and $\lambda_i$. Here we used the fact that $i_1^\bEB = i_1^\bV-3$ and $i_2^\bEB = i_2^\bV-2i_1^\bV+3$. Although $\mathcal{W}$ and $\overline{\mathcal{W}}$ are formally different energy functions with distinct dependencies, in this work we use $\mathcal{W}$ indiscriminately, as an energy density with dependencies implicitly defined by the invariants appearing in derivatives. 

For an isotropic material, the Bell stress~\cite{beatty1992deformations,steigmann2002invariants}, which is conjugate to the Bell strain, is given by
\begin{eqnarray}
  \boldsymbol\Sigma_{\text{Bell}}
  &=& \frac{\partial\mathcal{W}}{\partial\bEBl} =
      \bigg(\frac{\partial\mathcal{W}}{\partial i_1^{\textbf{E}_\text{B}}}
      +i_1^{\textbf{E}_\text{B}}\frac{\partial\mathcal{W}}{\partial i_2^{\textbf{E}_\text{B}}}
      +i_2^{\textbf{E}_\text{B}}\frac{\partial\mathcal{W}}{\partial i_3^{\textbf{E}_\text{B}}}\bigg)\mathbf{I}
      -\bigg(\frac{\partial\mathcal{W}}{\partial i_2^{\textbf{E}_\text{B}}}
      +i_1^{\textbf{E}_\text{B}}\frac{\partial\mathcal{W}}{\partial i_3^{\textbf{E}_\text{B}}} \bigg)\bEBl
      +\frac{\partial\mathcal{W}}{\partial i_3^{\textbf{E}_\text{B}}}\bEBl^2 \,.
\end{eqnarray}
The Bell stress is related to the Cauchy stress tensor through the relation
\begin{equation}
  \mathbf{T} = J^{-1}\mathbf{V}\cdot\boldsymbol\Sigma_{\text{Bell}}\,,
  \label{eq:rcauchy}
\end{equation}
where $J = i_3^{\mathbf{V}}$. If the material is incompressible, we have $J=1$, which can be factored into the variational principle presented in~\cite{vitral2022quadratic} through a constraint $p(J-1)$ in the energy, where the Lagrange multiplier $p$ is the pressure. In this case, the Bell stress derived from the Euler-Lagrange equations has the form
\begin{equation}
  \boldsymbol\Sigma_{\text{Bell}} = -p\bV^{-1}+\frac{\partial\mathcal{W}}{\partial\bEBl}
  = -p\bV^{-1}+
  \bigg(\frac{\partial\mathcal{W}}{\partial i_1^{\textbf{E}_\text{B}}}
  +i_1^{\textbf{E}_\text{B}}\frac{\partial\mathcal{W}}{\partial i_2^{\textbf{E}_\text{B}}}
  \bigg)\mathbf{I}
  -\frac{\partial\mathcal{W}}{\partial i_2^{\textbf{E}_\text{B}}}\bEBl \,.
  \label{eq:ibell}
\end{equation}
Therefore, for an incompressible quadratic-Biot material, the Bell~\eqref{eq:ibell} and Cauchy~\eqref{eq:rcauchy} stresses are
\begin{eqnarray}
  \boldsymbol\Sigma_{\text{Bell}} &=& -p\,\mathbf{V}^{-1}+[(2c_1+c_2)i_1^\bEB+c_2]\mathbf{I}-c_2\mathbf{V}\;,
  \label{eq:bst}
  \\[3mm]
  \mathbf{T} &=& -p\,\mathbf{I}+[(2c_1+c_2)i_1^\bEB+c_2]\mathbf{V}-c_2\bB\,.
  \label{eq:cauchy}
\end{eqnarray}

While it is straightforward to work with $\mathbf{B}$ and its invariants if the deformation is known, the same is not true for $\mathbf{V} = \sqrt{\mathbf{B}}$. An useful form of the stretch can be obtained through the Cayley-Hamilton theorem, which gives an explicit expression for $\mathbf{V}$ as a function of its own invariants and $\bB$~\cite{vitral2022quadratic,ting1985determination},
\begin{eqnarray}
  \mathbf{V} = \left( i_1^{\textbf{V}}i_2^{\textbf{V}}-i_3^{\textbf{V}} \right)^{-1}
  \left( i_1^{\textbf{V}}i_3^{\textbf{V}}\mathbf{I}+\left[ \left( i_1^{\textbf{V}} \right)^2
  -i_2^{\textbf{V}} \right]\bB-\bB^2 \right)  \, .
  \label{eq:vexp0}
\end{eqnarray}
This explicit form of $\bV$ can be substituted into the stresses~\eqref{eq:bst} and \eqref{eq:cauchy} in order to connect with usual expressions function of the tensor $\bB$.

\subsection{Uniaxial tension and linear limit}

\begin{figure}[ht]
  \centering
  \includegraphics[width=0.5\textwidth]{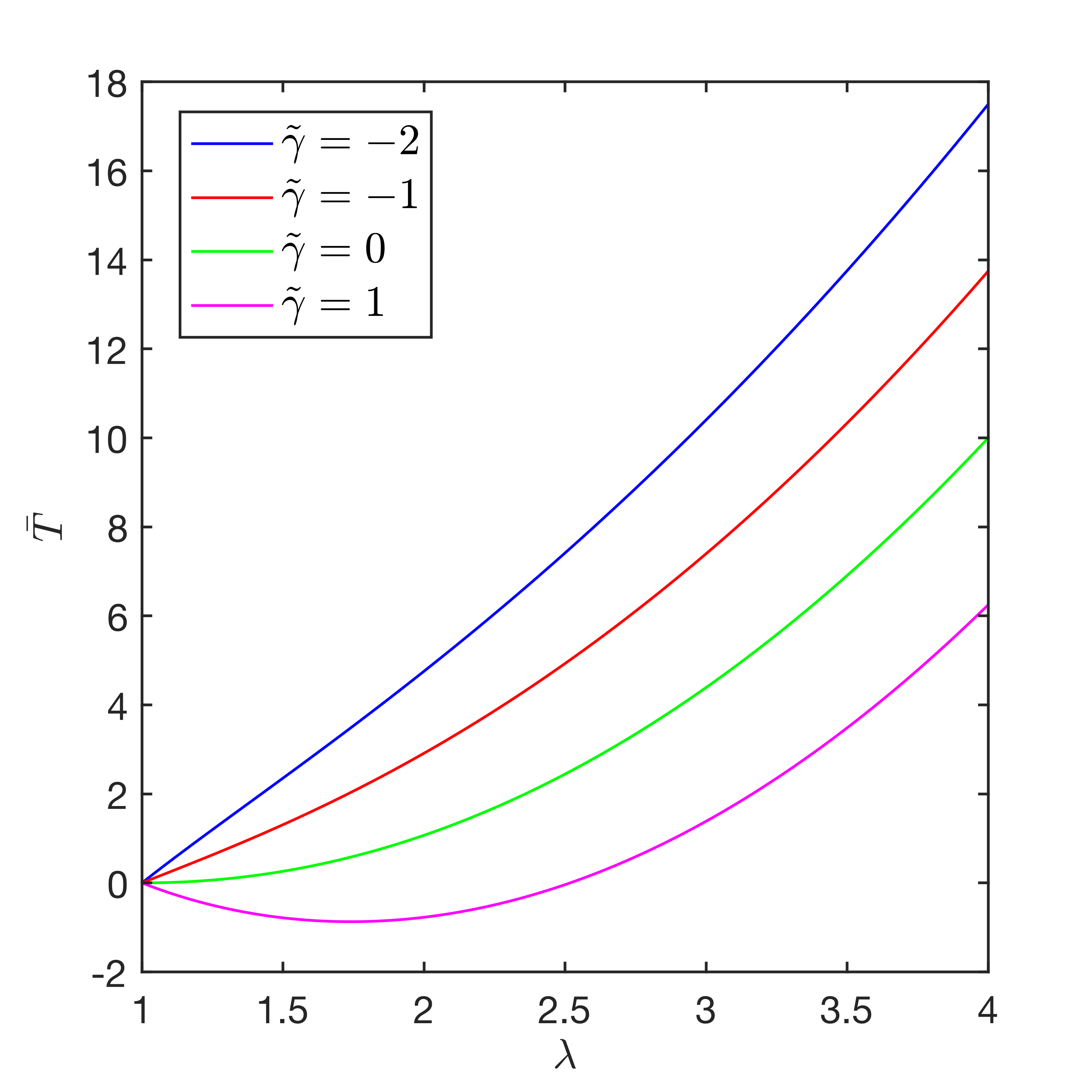}
  \caption{Nondimensional normal stress $\bar{T}$ for a quadratic-Biot material in uniaxial extension as a function of the stretch $\lambda$. The range of $\tilde\gamma = c_2/2c_1$ for which~\eqref{eq:wbiot} is convex is $-3/2 \leq \tilde\gamma \leq 0$.}
  \label{fig:ext}
\end{figure}

For familiarizing with the mechanical behavior of the incompressible quadratic-Biot material, it is insightful to evaluate its response with respect to a basic homogeneous deformation of uniaxial extension in the 1-direction. In this case, the principal stretches are $\lambda_1 = 1$, and $\lambda_2 = \lambda_3 = \lambda^{-1/2}$, so that
\begin{equation}
  \begin{aligned}
    \mathbf{V} &= \lambda\mathbf{e}_1\otimes\mathbf{e}_1 + \lambda^{-1/2}(\mathbf{e}_2\otimes\mathbf{e}_2+\mathbf{e}_3\otimes\mathbf{e}_3)\,,
    \\[2mm]
    \bB &= \lambda^2\mathbf{e}_1\otimes\mathbf{e}_1 + \lambda^{-1}(\mathbf{e}_2\otimes\mathbf{e}_2+\mathbf{e}_3\otimes\mathbf{e}_3)\,.
  \end{aligned}
  \label{eq:kine}
\end{equation}
\\
By assuming $\mathbf{T}\cdot\mathbf{e}_2 = \mathbf{T}\cdot\mathbf{e}_3 = \mathbf{0}$, we obtain the pressure $p$ from~\eqref{eq:cauchy}. We then substitute $p$ into the axial stress $T = \mathbf{T}\cdot\mathbf{e}_1$ and find
\begin{equation}
  T = [(2c_1+c_2)(\lambda+2\lambda^{-1/2}-3)+c_2](\lambda-\lambda^{-1/2})
  -c_2(\lambda^2-\lambda^{-1})\,.
\end{equation}
One convenient way to nondimensionalize the axial stress is
\begin{equation}
  \bar{T} = \frac{T}{2c_1} = [(1+\tilde\gamma)(\lambda+2\lambda^{-1/2}-3)
  +\tilde\gamma](\lambda-\lambda^{-1/2})-\tilde\gamma(\lambda^2-\lambda^{-1}) \,,
  \label{eq:ext}
\end{equation}
where $\tilde\gamma = c_2/2c_1$. The nondimensional axial stress $\bar{T}$ as a function of the stretch $\lambda$ is shown in Fig.~\ref{fig:ext}, for a range of $\tilde\gamma$ both negative and positive. The interval for which~\eqref{eq:wbiot} is convex is $-3/2 \leq \tilde\gamma \leq 0$. Observe that for $\tilde\gamma > 0 $, i.e. $c_2 > 0$, $\bar T$ initially decreases with $\lambda$, which is not physically reasonable.  For $\tilde\gamma < -3/2$, i.e. $c_1 < -c_2/3$, the behavior of the $\bar T$ curve becomes closer to linear (see $\tilde\gamma = -2$), which is physically reasonable for uniaxial tension, but lies outside the convexity range for~\eqref{eq:wbiot}.

  It is also helpful to connect the constants from the energy~\eqref{eq:wbiot} with the Lam\'e parameters in the linear limit. In linear elasticity, the isotropic Cauchy stress is given by $\mathbf{T} = \lambda_L (\textrm{Tr}\,\boldsymbol\varepsilon)\mathbf{I}+2\mu_L\boldsymbol\varepsilon$, where $\lambda_L$ and $\mu_L$ are the first and second Lam\'e parameters, respectively, and $\boldsymbol\varepsilon$ is the small strain tensor. By linearizing $\bV \approx \mathbf{I} +\boldsymbol\varepsilon$ and $\bB \approx \mathbf{I} +2\boldsymbol\varepsilon$, from the Cauchy stress for the incompressible quadratic-Biot material~\eqref{eq:cauchy} we find that
\begin{equation}
  c_2 = -2\mu_L\,.
\end{equation}
Therefore, the condition $c_2 \leq 0$ implies the usual $\mu_L \geq 0$. The relation $\lambda_L = 2c_1+c_2$ can be identified from the Cauchy stress when incompressibility is not enforced for the material~\eqref{eq:wbiot}. It is also straightforward to identify these relations by rewriting the energy~\eqref{eq:wbiot} as~\cite{vitral2022quadratic}
\begin{equation}
  \mathcal{W} = \bigg(c_1+\frac{c_2}{2}\bigg)\textrm{Tr}^2\,\bEB -\frac{c_2}{2}\,\textrm{Tr}(\bEB^2)\,.
\end{equation}

\section{Simple shear}
\label{sec:shear}

We now formulate an incompressible simple shear deformation in terms of stretches and Bell strains, adopting the quadratic-Biot material as the main example, and contrast this formulation with the classic one on the basis of Cauchy-Green tensors. Under traction free lateral boundaries, generalized neo-Hookean materials do not present the Poynting effect for such deformation, which highlights the importance of the second invariant for materials governed by the invariants of $\bB$. Moreover, neo-Hookean and Mooney-Rivlin materials present no shear hardening under simple shear. While many materials show a linear relation between shear stress and amount of shear, these two models miss other important nonlinear effects, requiring additional terms in the energy to be recovered~\cite{mooney1940theory,mangan2016strain}. This section aims to clarify how these observations compare with the mechanical response of a quadratic-Biot material and understand the role played by the invariants of the tensors $\bV$ and $\bEBl$ on stretch based formulations.

The isochoric deformation for simple shear is given by~\cite{horgan2010simple}
\begin{equation}
  x_1 = X_1+\kappa X_2\,,\quad x_2 = X_2\,,\quad x_3 = X_3 \,,
  \label{eq:simple}
\end{equation}
where $X_i$ are reference coordinates of $\mathbf{X}$, $x_i$ deformed coordinates of $\mathbf{x}$, and $\kappa$ is the amount of shear ($\kappa = \textrm{tan}\,\theta$, for a shear angle $\theta$). We adopt rectangular Cartesian coordinates, with a basis $\{\mathbf{e}_i\}$. Traction free boundary condition is assumed in the out-of-plane direction, $\mathbf{T}\cdot\mathbf{e}_3 = \mathbf{0}$. Note that in experiments deformations such as~\eqref{eq:simple} are hard to be controlled, see Destarde et al.~\cite{destrade2012simple} for further discussion.

Based on~\eqref{eq:simple}, the deformation gradient $\mathbf{F}$, the left Cauchy-Green deformation tensor $\bB$ and the squared tensor $\mathbf{B}^2$ have the following form
\begin{equation}
  \begin{aligned}
    \mathbf{F} &= \mathbf{I} + \kappa\mathbf{e}_1\otimes\mathbf{e}_2\,,
    \\[2mm]
    \bB &= \mathbf{I} +\kappa^2\mathbf{e}_1\otimes\mathbf{e}_1+\kappa(\mathbf{e}_1\otimes\mathbf{e}_2+\mathbf{e}_2\otimes\mathbf{e}_1)\,,
    \\[2mm]
    \bB^2 &= \mathbf{I}+(\kappa^4+3\kappa^2)\mathbf{e}_1\otimes\mathbf{e}_1
    +\kappa^2\mathbf{e}_2\otimes\mathbf{e}_2
    +(\kappa^3+2\kappa)(\mathbf{e}_1\otimes\mathbf{e}_2+\mathbf{e}_2\otimes\mathbf{e}_1)\,.
  \end{aligned}
  \label{eq:kine}
\end{equation}

From~\cite{vitral2022quadratic}, we can find exactly the first invariant of $\bEBl$ for simple shear as $i_1^\bEB = -2+\sqrt{i_1^{\bB}+2J-1} = -2 +\eta$, where $\eta = \sqrt{4+\kappa^2}$. Consequently, $i_1^{\mathbf{V}} = i_2^{\mathbf{V}} = 1+\eta$, and $i_2^\bEB = 1-\eta$. Instead of computing the left stretch through $\mathbf{V} = \sqrt{\bB}$, we substitute these invariants into the explicit expression for $\mathbf{V}$~\eqref{eq:vexp0}, and obtain
\begin{equation}
    \mathbf{V} = \frac{(\eta+1)\mathbf{I}+(\eta^2+\eta)\bB-\bB^2}{\eta^2+2\eta}\,.
    \label{eq:vexp}
\end{equation}

By substituting the previous expression for $\mathbf{V}$ into~\eqref{eq:cauchy}, we find the following components of the Cauchy stress
\begin{equation}
  \begin{aligned}
  &T_{11} = -p +\bigg[\frac{\partial\mathcal{W}}{\partial i_1^\bEB}+(i_1^\bEB+1)\frac{\partial\mathcal{W}}{\partial i_2^\bEB}\bigg]\bigg[1+\frac{\kappa^2(\eta+1)}{\eta^2+2\eta}\bigg]-\frac{\partial\mathcal{W}}{\partial i_2^\bEB}(\kappa^2+1)\,,
  \\[2mm]
  &T_{22} = -p + \bigg[\frac{\partial\mathcal{W}}{\partial i_1^\bEB}+(i_1^\bEB+1)\frac{\partial\mathcal{W}}{\partial i_2^\bEB}\bigg]\bigg(1-\frac{\kappa^2}{\eta^2+2\eta}\bigg)-\frac{\partial\mathcal{W}}{\partial i_2^\bEB}\,,
  \\[2mm]
  &T_{33} = -p + \frac{\partial\mathcal{W}}{\partial i_1^\bEB}+i_1^\bEB\frac{\partial\mathcal{W}}{\partial i_2^\bEB}\,,
  \\[2mm]
  &T_{12} = \bigg[\frac{\partial\mathcal{W}}{\partial i_1^\bEB}+(1+i_1^\bEB)\frac{\partial\mathcal{W}}{\partial i_2^\bEB}\bigg]\frac{\kappa}{\eta}-\frac{\partial\mathcal{W}}{\partial i_2^\bEB}\kappa\,.
  \label{eq:tcomp}
  \end{aligned}
\end{equation}
As characteristic of stretch-based formulations, these expressions are not as simple as their counterparts in terms of invariants of $\mathbf{B}$~\cite{rivlin1948large,wineman2005some}. Nevertheless, of course Rivlin universal relation $T_{11}-T_{22} = \kappa T_{12}$ for the displacement formulation~\eqref{eq:simple} holds, which implies that normal stresses are required to maintain the shear stress $T_{12}$. For completeness, we can also express these components on the basis of $i_i^\bV$ ,
\begin{equation}
\begin{aligned}
  &T_{11} = -p +\bigg(\frac{\partial\mathcal{W}}{\partial i_1^\bV}+i_1^\bV\frac{\partial\mathcal{W}}{\partial i_2^\bV}\bigg)\bigg[1+\frac{\kappa^2(\eta+1)}{\eta^2+2\eta}\bigg]-\frac{\partial\mathcal{W}}{\partial i_2^\bV}(\kappa^2+1)\,,
  \\[2mm]
  &T_{22} = -p + \bigg(\frac{\partial\mathcal{W}}{\partial i_1^\bV}+i_1^\bV\frac{\partial\mathcal{W}}{\partial i_2^\bV}\bigg)\bigg(1-\frac{\kappa^2}{\eta^2+2\eta}\bigg)-\frac{\partial\mathcal{W}}{\partial i_2^\bV} \,,
  \\[2mm]
  &T_{33} = -p + \frac{\partial\mathcal{W}}{\partial i_1^\bV}+(i_1^\bV-1)\frac{\partial\mathcal{W}}{\partial i_2^\bV}\,,
  \\[2mm]
  &T_{12} = \bigg(\frac{\partial\mathcal{W}}{\partial i_1^\bV}+i_1^\bV\frac{\partial\mathcal{W}}{\partial i_2^\bV}\bigg)\frac{\kappa}{\eta}-\frac{\partial\mathcal{W}}{\partial i_2^\bV}\kappa\,.
  \label{eq:compv}
\end{aligned}
\end{equation}

Different approaches exist for determining the pressure, including plane stress and zero normal traction formulations (a detailed discussion can be found in Horgan and Murphy~\cite{horgan2010simple}). Here we adopt the former, through which $p$ can be obtained from the out-of-plane traction boundary condition $\mathbf{T}\cdot\mathbf{e}_3=  \mathbf{0}$, that is, $T_{33} = 0$. In this case, the remaining normal stress components become
\begin{eqnarray}
  \nonumber
  T_{11} &=& \frac{\kappa^2(\eta+1)}{\eta^2+2\eta}\bigg[\frac{\partial\mathcal{W}}{\partial i_1^\bEB}+(i_1^\bEB+1)\frac{\partial\mathcal{W}}{\partial i_2^\bEB}\bigg]-\frac{\partial\mathcal{W}}{\partial i_2^\bEB}\kappa^2 = \frac{\kappa^2(\eta+1)}{\eta^2+2\eta}\bigg(\frac{\partial\mathcal{W}}{\partial i_1^\bV}+i_1^\bV\frac{\partial\mathcal{W}}{\partial i_2^\bV}\bigg)-\frac{\partial\mathcal{W}}{\partial i_2^\bV}\kappa^2\,,
  \\[2mm]
  T_{22} &=& \frac{-\kappa^2}{\eta^2+2\eta}\bigg[\frac{\partial\mathcal{W}}{\partial i_1^\bEB}+(i_1^\bEB+1)\frac{\partial\mathcal{W}}{\partial i_2^\bEB}\bigg] = \frac{-\kappa^2}{\eta^2+2\eta}\bigg(\frac{\partial\mathcal{W}}{\partial i_1^\bV}+i_1^\bV\frac{\partial\mathcal{W}}{\partial i_2^\bV}\bigg)\,.
             \label{eq:normal}
\end{eqnarray}
We can now compare the normal stress~\eqref{eq:normal} and the shear stress~\eqref{eq:compv} with their counterparts~\cite{horgan2010simple} in terms of invariants of $\bB$, also assuming $T_{33} = 0$, which present a much simpler form
\begin{align}
  T_{11} = 2\kappa^2\frac{\partial\mathcal{W}}{\partial i_1^\bB}\,, \quad
  T_{22} = -2\kappa^2\frac{\partial\mathcal{W}}{\partial i_2^\bB}\,, \quad
  T_{12} = 2\kappa\bigg(\frac{\partial\mathcal{W}}{\partial i_1^\bB}+\frac{\partial\mathcal{W}}{\partial i_2^\bB}\bigg)\,.
  \label{eq:compb}
\end{align}

As an example, we will contrast the mechanical behavior of a quadratic-Biot material~\eqref{eq:wbiot} with neo-Hookean and Mooney-Rivlin materials 
\begin{align}
  \label{eq:nh}
  &\mathcal{W}_{nH} = c_{nH} \,(i_1^\bB-3)\,, \\[2mm]
  &\mathcal{W}_{MR} = c^{MR}_1\,(i_1^\bB-3)+c^{MR}_2\,(i_2^\bB-3)\,,
  \label{eq:mr}
\end{align}
and other incompressible material laws based on the invariants of $\bB$. Based on~\eqref{eq:cauchy} and \eqref{eq:vexp}, we can write the Cauchy stress for the incompressible quadratic-Biot material as
\begin{eqnarray}
  \mathbf{T} &=& [-p+\zeta(\eta+1)]\mathbf{I} + [\zeta(\eta^2+\eta)-c_2]\bB-\zeta\bB^2\,,
                 \quad\textrm{where}\quad
                 \zeta = \frac{(2c_1+c_2)i_1^\bEB+c_2}{\eta^2+2\eta} \,.
    \label{eq:scauchy}
\end{eqnarray}
The pressure $p$ can be determined from the out-of-plane boundary condition $T_{33} = 0$, which gives
\begin{equation}
  p = (2c_1+c_2)i_1^\bEB\,.
\end{equation}
Since $\kappa$ is constant, the pressure is also constant throughout the body, so that the equilibrium equations for the homogeneous deformation~\eqref{eq:simple} are satisfied when body forces are zero.

Before discussing the stress components, it is important to evaluate whether and under which circumstances the response functions in~\eqref{eq:scauchy} satisfy commonly considered inequalities in solid mechanics.

\subsection{Restrictions on response functions}
\label{sec:rest}

Since the response functions of a stretch based constitutive description of~\eqref{eq:cauchy} are not conventionally found in the literature, it is important to know what form takes the traditional response functions $\beta_i$ for a quadratic-Biot material. These assist in determining if adscititious inequalities that suggest physically realistic deformations are satisfied. That is, when writing the Cauchy stress tensor of as
\begin{equation}
  \mathbf{T} = \beta_0\mathbf{I}+ \beta_1\bB+\beta_{-1}\bB^{-1}
\end{equation}
it is often assumed that $\beta_i$ satisfy certain restrictions in order to represent the physical behavior of hyperelastic materials~\cite{beatty1987topics}. Among them are the Baker-Ericksen (BE) and the empirical (E) inequalities: the former follows from observations that the largest principal stress lies in the direction of the largest principal stretch~\cite{baker1954inequalities}, and the latter consists of stronger restrictions postulated by Truesdell~\cite{truesdell1952mechanical,truesdell2004non} based on available experimental evidence at the time. The E inequalities are given by $\beta_0 \leq 0$, $\beta_1 > 0$ and $\beta_{-1} \leq 0$, or simply $\beta_1 > 0$ and $\beta_{-1} \leq 0$ in the incompressible case, which are known to hold for many rubber-like materials. However, due to its lack of theoretical foundation the E inequalities have been fairly criticized for arbitrarily restricting hyperelastic energy densities~\cite{mihai2011positive,liu2012note,thiel2019we} and not actually preventing physically unrealistic responses~\cite{saravanan2011adequacy}. In particular, for capturing the reverse Poynting effect in simple shear, the condition $\beta_{-1} \leq 0$ is necessarily violated~\cite{mihai2011positive}. Experiments using a rheometer for shearing bio-gels suggest the existence of such a reverse effect in torsion~\cite{janmey2007negative}; however, their microstructure and macroscopic behavior differ from elastomer type materials, so that the reverse effect could also be explained by the anisotropy of the material~\cite{destrade2015dominant}, as is the case of soft composites~\cite{araujo2020experimental}.

By the Cayley-Hamilton theorem, it can be shown from~\eqref{eq:scauchy} that the response function under simple shear for an incompressible quadratic-Biot material is
\begin{equation}
  \beta_1 = \frac{(\eta+1)(2c_1+c_2)i_1^\bEB-c_2}{\eta^2+2\eta}\,,\quad
  \beta_{-1} = -\frac{(2c_1+c_2)i_1^\bEB+c_2}{\eta^2+2\eta}\,.
\end{equation}
For a simple shear deformation~\eqref{eq:simple}, as discussed in~\cite{mihai2011positive} the BE inequalities hold if and only if $\beta_1 > \beta_{-1}$, which requires a stronger condition on the quadratic-Biot material constants to be true: $c_1 > -c_2/2$. Note that it does not restrict the sign of $c_2$. In case $c_2 \leq 0$, the E inequality $\beta_1 > 0$ is satisfied for any $\kappa$, but the response function $\beta_{-1}$ can be both negative or positive depending on the values of $c_1$ and $c_2$, and even change sign as a function of $\kappa$, so that the second E inequality does not hold. The generalized empirical inequalities, proposed by Mihai and Goriely~\cite{mihai2013numerical}, solve this issue by relaxing the condition on $\beta_{-1}$. If we allow $c_2 > 0$, then we can have a scenario where the second E inequality is satisfied, but not the first one. It is also possible to evaluate restrictions on material parameters based on a thermodynamic stability analysis, which is shown by Liu~\cite{liu2012note} to give less restrictive conditions than the E inequalities for uniaxial contraction. 

\subsection{Poynting effect and shear hardening}

We proceed with an investigation of shear hardening in a quadratic-Biot material. By substituting the energy~\eqref{eq:wbiot} into the shear stress $T_{12}$ from~\eqref{eq:tcomp}, we obtain
\begin{equation}
  T_{12} = \frac{(2c_1+c_2)\kappa}{\eta}\,i_1^\bEB-c_2\kappa\bigg(1-\frac{1}{\eta}\bigg)
  =  2c_1 \kappa\bigg(1-\frac{2}{\eta}\bigg)-c_2\frac{\kappa}{\eta}\,.
  \label{eq:t12}
\end{equation}
Observe that in the limit of small $\kappa$, the shear stress presents a linear relation $T_{12} \sim -(c_2/2)\kappa$, similarly to neo-Hookean and Mooney-Rivlin materials -- see~\eqref{eq:compb}. In the less realistic limit of large $\kappa$, we also approach a linear relation $T_{12} \sim 2c_1\kappa-c_2$. In between, the shear behavior is clearly nonlinear with a derivative
\begin{equation}
  \frac{\textrm{d} T_{12}}{\textrm{d} \kappa} = 2c_1-\frac{16c_1+4c_2}{\eta^3}\,.
\end{equation}

\begin{figure}[ht]
  \centering
  \includegraphics[width=0.5\textwidth]{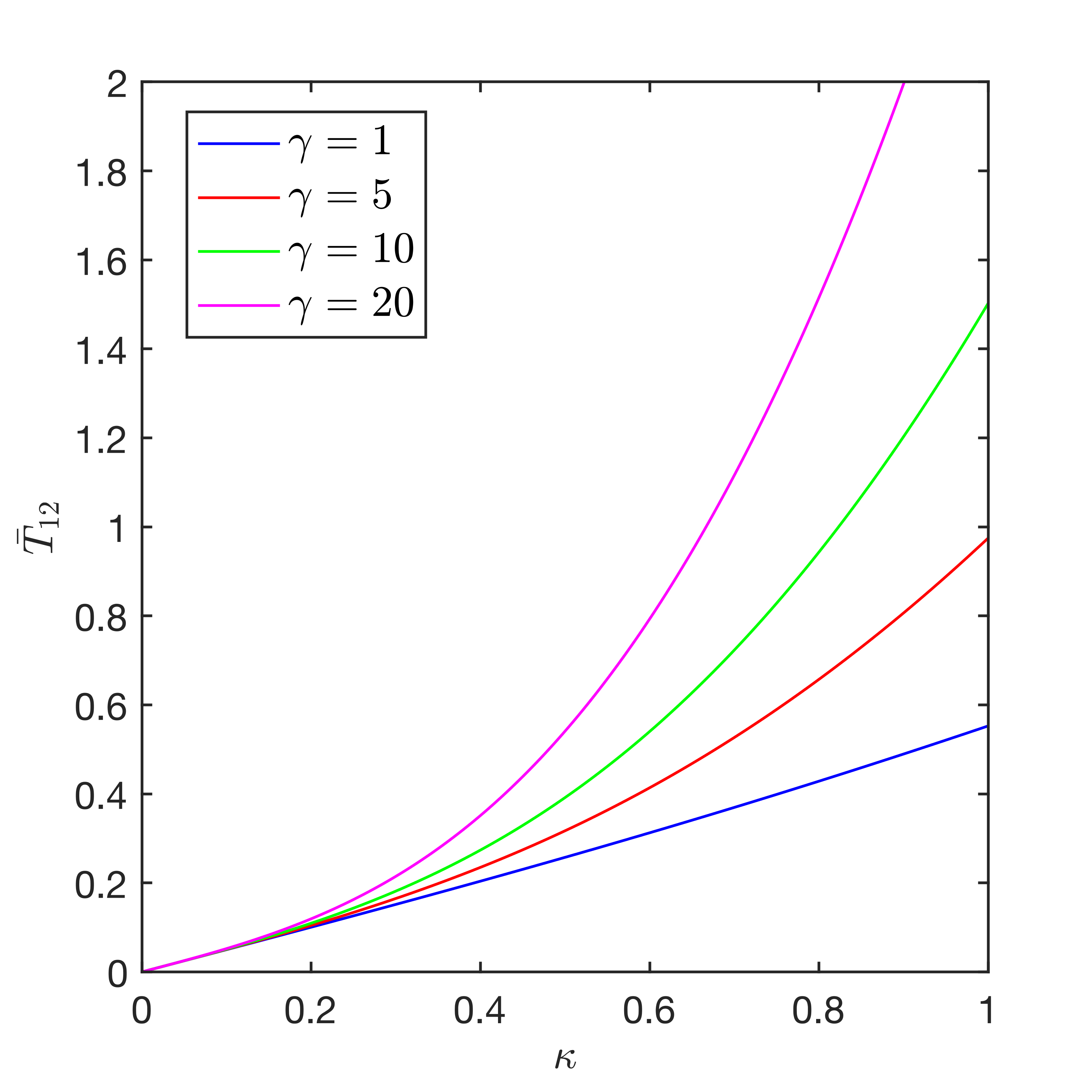}
  \caption{Nondimensional shear stress $\bar{T}_{12}$ for a quadratic-Biot material in simple shear as a function of the amount of shear $\kappa$. Shear hardening increases with the ratio $\gamma = -2c_1/c_2$ between the material's constants.}
  \label{fig:t12}
\end{figure}

One way to nondimensionalize the shear stress $T_{12}$~\eqref{eq:t12} is by dividing it by $-c_2$,
\begin{align}
  &\bar{T}_{12} = \frac{T_{12}}{-c_2} = \frac{\kappa(\gamma-1)}{\eta}i_1^\bEB+\kappa\bigg(1-\frac{1}{\eta}\bigg)\,,
\end{align}
where $\gamma = -2c_1/c_2 > 1$ since $c_1 > -c_2/2$ from the discussion in Section~\ref{sec:rest}. In Fig.~\ref{fig:t12} we plot $\bar{T}_{12}$ as a function of the amount of shear for different values of $\gamma$. For the limiting case $\gamma = 1$ the curve is approximately linear up to $\kappa=1$, whereas as $\gamma$ increases, a nonlinear shear hardening response intensifies. No shear softening is observed for the allowable values of $\gamma$ -- this would require the violation of the convexity condition $c_1 \geq -c_2/3$. This is a deficiency of the quadratic-Biot material, since shear softening has been experimentally observed for incompressible solids~\cite{nunes2013simple}, and is an effect captured by generalized neo-Hookean models~\cite{anssari2021modelling,anssari2022three}.

Interestingly, from~\eqref{eq:compb} we see that incompressible materials with an energy linear in $\bB$ invariants (e.g. Mooney-Rivlin) will not present shear hardening or softening, whereas for the stretch based formulations~\eqref{eq:tcomp} and \eqref{eq:compv} we see that even energies linear in $\bV$ invariants (e.g. generalized Varga) will display a nonlinear $T_{12}$ in $\kappa$. This is not a deficiency of Cauchy-Green formulations, but a comparison between different functional forms linear in a different set of invariants. Another example is the one-term Ogden material~\cite{ogden1972large}, function of $(\lambda_1^n+\lambda_2^n+\lambda_3^n-3)$, for which the shear stress $T_{12}$ is nonlinear in $\kappa$, except for $n = \pm 2$.

Another way to nondimensionalize $T_{12}$ would be to divide~\eqref{eq:t12} by $2c_1$, so that
\begin{equation}
  \frac{T_{12}}{2c_1} =  \kappa\bigg(1-\frac{2}{\eta}\bigg)-\tilde\gamma\frac{\kappa}{\eta}\,.
\end{equation}
Due to the BE inequalities, $\tilde\gamma = c_2/2c_1 > -1$. If $\tilde\gamma > 0$, i.e. $c_2 > 0$, then the response is a nondimensional shear stress that initially decays with $\kappa$ towards negative values, similar to the response for $\tilde\gamma = 1$ observed for uniaxial extension in Fig.~\ref{fig:ext}, which is not physically reasonable. Hence, we restrict $\tilde\gamma$ to the range $0 \geq \tilde\gamma > -1$.

In order to evaluate the presence and nature of the Poynting effect in a quadratic-Biot material, we calculate from~\eqref{eq:scauchy} the stress component in the normal direction to the applied shear,
\begin{equation}
    T_{22} =  \frac{-\kappa^2}{\eta^2+2\eta}[(2c_1+c_2)i_1^\bEB+c_2]\,.
    \label{eq:poy}
\end{equation}
Recall that we restrict $c_1 > -c_2/2$ and $c_2 \leq 0$. We conclude that for $c_2 = 0$ a quadratic-Biot material presents the classic Poynting effect, that is, $T_{22} < 0$. However, at least for small $\kappa$, the reverse Poynting effect, $T_{22} > 0$, may be observed when $c_2 < 0$, so that $i_2^\bEB$ plays a role in switching the effect from positive to negative normal stress. Note that the expression $(2c_1+c_2)i_1^\bEB+c_2$ is exactly the coefficient of $\mathbf{V}$ in the Cauchy stress~\eqref{eq:cauchy}. This term is absent in the $\mathbf{T}$ expression for a generalized neo-Hookean material, which only presents the pressure and $\bB$ terms, and hence the Poynting effect is not observed. While we focus on isotropic materials, the nature of $T_{22}$ may also depend on anisotropy, as analyzed by Horgan and Murphy~\cite{horgan2017poynting} in the context of soft fibrous materials.

\begin{figure}[ht]
  \centering
  \includegraphics[width=0.5\textwidth]{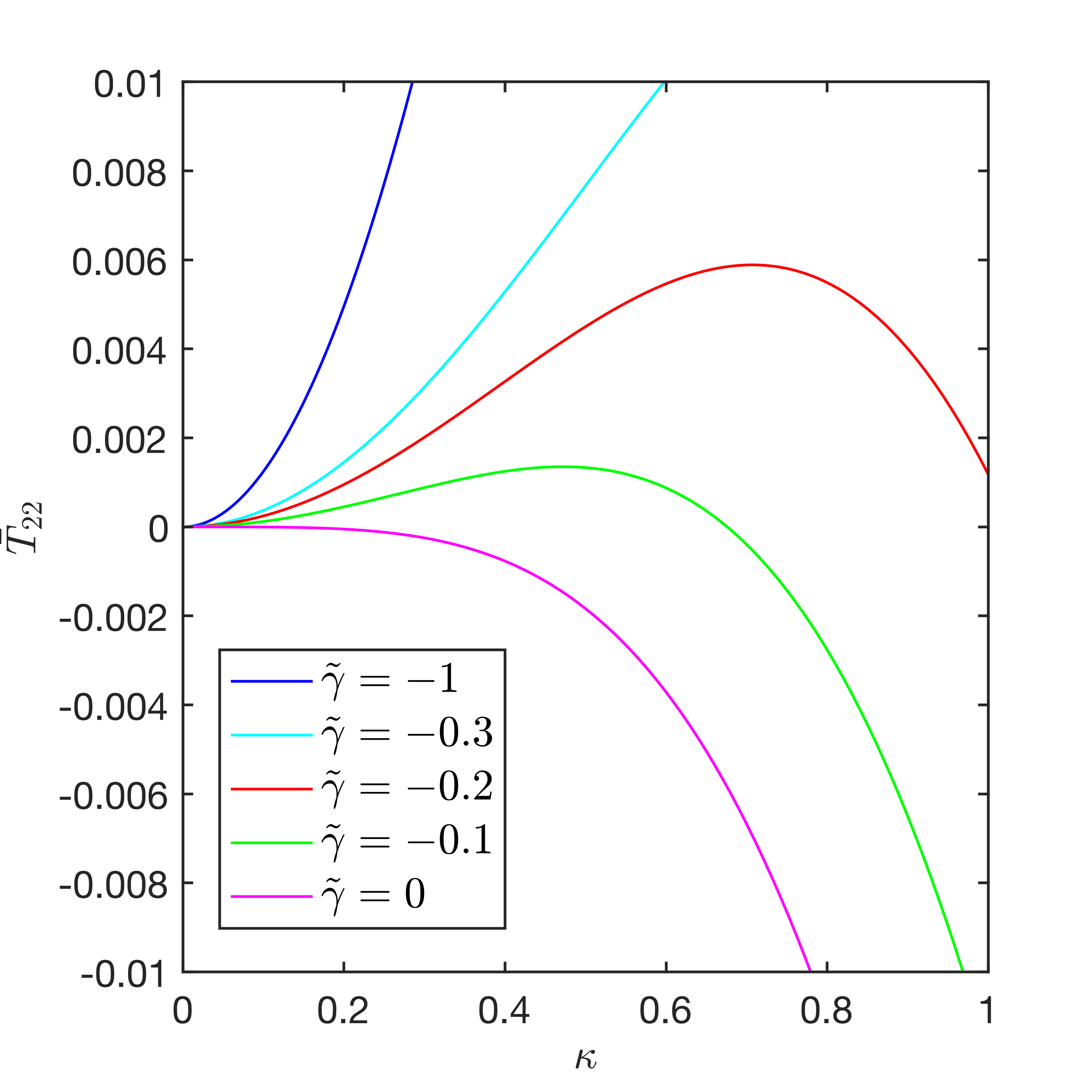}
  \caption{Nondimensional normal stress $\bar{T}_{22}$ for a quadratic-Biot material in simple shear as a function of the amount of shear $\kappa$. Observe a transition from the classic Poynting effect ($\bar{T}_{22}<0$) to the reverse Poynting effect ($\bar{T}_{22} > 0$) as the ratio $\tilde{\gamma} = c_2/2c_1$ decreases from 0 towards -1.}
  \label{fig:t22}
\end{figure}

By dividing the normal stress $T_{22}$ by $2c_1$, we obtain the nondimensional quantity
\begin{equation}
  \bar{T}_{22} = \frac{T_{22}}{2c_1} = \frac{-\kappa^2}{\eta^2+2\eta}[(1+\tilde{\gamma})i_1^\bEB+\tilde{\gamma}]\,,
  \label{eq:poy2}
\end{equation}
where $0 \geq \tilde{\gamma} = c_2/2c_1 > -1$. In Fig.~\ref{fig:t22} we plot a family of $\bar{T}_{22}$ curves as a function of $\kappa$ for different values of $\tilde{\gamma}$. When $\tilde{\gamma} = 0$, only the classic Poynting effect is observed. By decreasing $\tilde{\gamma}$, the normal stress starts as $\bar{T}_{22} > 0 $ and transits to $\bar{T}_{22} < 0$ as $\kappa$ increases, displaying a downward concavity maximum with $\bar{T}_{22}$ at a critical amount of shear $\kappa_c = [-2\tilde{\gamma}/(1+\tilde{\gamma})]^{1/2}$. This normal strain softening is reminiscent of the Mullins effect in rubbers~\cite{dorfmann2004constitutive}. Such transition is suggested, for example, from the data of Janmey et al.~\cite{janmey2007negative} presented by Destrade et al.~\cite{destrade2015dominant} for the shearing of a block of gel made from actin cross-linked by polyacrylamide, which shows an initial small region of reverse Poynting before switching to classic Poynting as $\kappa$ increases. For $\tilde{\gamma} \leq -0.3$ only the reverse Poynting effect is observed for physically reasonable values of $\kappa$. The ratio between material's constants clearly determines which nature of Poynting is the dominant one, so that, for a fixed $c_1$, by decreasing $c_2$ we can switch from classic to reverse Poynting effect.

\subsection{Discussion}

From the stress components~\eqref{eq:compb}, we see one example of why the second invariant of $\bB$ plays an important role in Cauchy-Green based modeling in nonlinear elasticity~\cite{wineman2005some,horgan2012importance,anssari2021central}. Although adding a dependence on $i_2^\bB$ to such an energy solves the ``incompleteness'' issue of generalized neo-Hookean materials $\mathcal{W}_1$, the lingering question is why a $\mathcal{W}_1$ material fails in modeling certain mechanical behaviors. Since $\lambda_ i = 1+\Delta_i$, principal stretches are of order one and both terms $(i_1^\bB-3)$ and $(i_2^\bB-3)$ are of order $O(\Delta_i)$ in principal strains. Hence, the addition of $i_2^\bB$ does not provide a next order contribution to $\mathcal{W}_1$ in a small strain expansion. The present result~\eqref{eq:poy} suggests that there is a more fundamental reason behind the the limitations of the $\mathcal{W}_1$ class, such as the absence of Poynting effect in simple shear under plane stress assumption.

While both neo-Hookean and quadratic-Biot materials are quadratic in stretch, the latter presents a more complete collection of stretches that appears from a systematic expansion in small Bell strains. For a particular energy, it does not matter which strain or deformation tensor is adopted to formulate the problem, since a set of invariants can be translated into another~\cite{hoger1984determination}. However, when comparing different energies that are limited to the first invariant of $\bB$, against those limited to the first invariant of $\bV$ or some other restricted combination of stretch powers, then differences in mechanical behavior may appear. In this case, simple shear is insightful for contrasting these various constructions. For example, compare the following incompressible materials: Varga, neo-Hookean, and quadratic-Biot materials. The Varga model~\cite{varga1966stress},
\begin{equation}
  \mathcal{W}_V = 2c_V\,i_1^\bEB\,,
  \label{eq:varga}
\end{equation}
is equivalent to a $n=1$ Ogden~\cite{ogden1972large} model, which is a formalization of the idea of modeling an incompressible material with independent principal stretches $\lambda_i$.

In this regard, the Varga~\eqref{eq:varga} and neo-Hookean~\eqref{eq:nh} models differ by the choice of strain employed in these one-constant energies: the former adopts the (first invariant of) Bell strain, and the latter the Green-Lagrange strain. By comparing~\eqref{eq:tcomp} and \eqref{eq:compb}, we see that this seemingly innocuous choice dictates whether the Poynting effect will be present or not for the problem under consideration. For the quadratic-Biot material, the classic Poynting effect appears even if $c_2 = 0$, that is, when the second invariant of $\bEBl$ (or $\mathbf{V}$) is not present, which is the case of the Varga model. Hence, a class of materials function only of the first invariant of $\bV$ or $\bEBl$, tensors linear in stretch, present significant differences in mechanical behavior when compared to the $\mathcal{W}_1$ class.

We next look into the functional form of $\mathcal{W}_1$ materials to understand how the second normal stress vanishes under plane stress assumption in simple shear. Since from~\eqref{eq:kine} we have $B_{22} = B_{33} = 1$, one can observe in $\mathbf{T}$~\eqref{eq:cauchy} that any contribution from the $\bB$ term to $T_{22}$ is eliminated by the pressure due to the traction free condition $T_{33} = 0$. From the explicit expression for $\mathbf{V}$~\eqref{eq:vexp}, we perceive that the only nonzero contribution for $T_{22}$ comes from $\bB^2$, since $(B^2)_{22} = 1+\kappa^2$ and $(B^2)_{33} = 1$. This contribution from $\bB^2$ is absent for generalized neo-Hookean materials, which becomes clear from the Cauchy-Green formulation of $\mathbf{T}$,
\begin{equation}
  \mathbf{T} =
     -p\,\mathbf{I}
      +2\bigg(\frac{\partial\mathcal{W}}{\partial i_1^{\bB}}
      +\frac{\partial\mathcal{W}}{\partial i_2^{\bB}}\,i_1^{\bB}\bigg)\bB
      -2\frac{\partial\mathcal{W}}{\partial i_2^{\bB}}\,\bB^2\,.
      \label{eq:cauchy-cg}
\end{equation}

From another angle, observe from $\bV$~\eqref{eq:vexp} that the explicit expression for the Bell strain in simple shear is
\begin{equation}
    \bEBl = \frac{(-\eta^2-\eta+1)\mathbf{I}+(\eta^2+\eta)\bB-\bB^2}{\eta^2+2\eta}\,,
\end{equation}
so that $\bEBl$ requires up to the quadratic power of $\bB$ in order to be represented by the left Cauchy-Green tensor. Although the Bell (or Biot) strain is the most primitive measure of strain one could define from a stretch, it cannot be found in the Cauchy stress tensor derived from an energy of the type $\mathcal{W}_1$, since $\mathbf{T}$~\eqref{eq:cauchy-cg} for this class does not present $\bB^2$. That is, some information about the Bell strain is lost for such materials, so that even for small nonlinear deformations they may fail to present a mechanical behavior that a linear stretch Varga model could display.

This reasoning is consistent with the observation about the importance of $i_2^{\bB}$ in nonlinear elasticity~\cite{horgan2012importance,anssari2021central}, since the coefficient of $\bB^2$ in the Cauchy stress~\eqref{eq:cauchy-cg} for an isotropic material becomes nonzero when $i_2^{\bB}$ appears in the energy. Hence, models such as Mooney-Rivlin provide a collection of stretches that can more accurately represent Bell strains (or, equivalently, the left stretch $\mathbf{V}$), and recover Poynting and other effects that are lost in the $\mathcal{W}_1$ class.

Even so, the addition of $i_2^{\bB}$ is not enough to reproduce all the responses displayed by a quadratic-Biot material. For example, from \eqref{eq:mr} and \eqref{eq:compb}, in simple shear the two constant Mooney-Rivlin material presents $T_{22} = -2c_2^{MR}\kappa^2$. When relaxing the E inequalities, a Mooney-Rivlin solid can show both compressive or tensile second normal stress, but not the local maximum and $T_{22}$ transition as a function of $\kappa$ discussed in \eqref{eq:poy2} and Fig.~\ref{fig:t22}. For further comparison, we can use explicit expressions for $\bB$ invariants as a function of $\bV$ invariants for an incompressible material~\cite{hoger1984determination}, and recast the Mooney-Rivlin and neo-Hookean energy as
\begin{align}
  &\mathcal{W}_{nH} = c_{nH} \,[(i_1^\bV)^2-2i_2^\bV-3]\,, \\[2mm]
  &\mathcal{W}_{MR} = c^{MR}_1(i_1^\bV)^2+c_2^{MR}(i_2^\bV)^2-2(c_1^{MR}\,i_2^\bV+c_2^{MR}i_1^\bV)-3(c_1^{MR}+c_2^{MR})\,.
\end{align}
Observe that the Mooney-Rivlin material is quadratic in both invariants of $\bV$, whereas the quadratic-Biot material~\eqref{eq:biotv} is only quadratic in $i_1^\bV$ and linear in $i_2^\bV$. In contrast to~\eqref{eq:biotv}, the neo-Hookean material lacks a linear $i_1^\bV$ term.

\section{Pure torsion}
\label{sec:torsion}

In this session, we briefly discuss the Poynting effect for the pure torsion of an isotropic incompressible solid cylinder. The earlier works of Rivlin~\cite{rivlin1948large,rivlin1949large6} on pure torsion of such cylinder adopted the usual energy function of invariants of the Cauchy-Green tensor. Later, Rivlin~\cite{rivlin2004note} reformulated the equations in terms of stretch invariants, which have been generalized by Horgan and Murphy~\cite{horgan2011extension} for the case of torsion superimposed on axial elongation. Here we show an equivalent formulation in terms of Bell strains and explore the difference in resultant axial force between the quadratic-Biot, neo-Hookean and Varga models.

In a cylindrical coordinate system, the isochoric pure torsion deformation (no radial stretch) of a solid cylinder of radius $a$ is
\begin{equation}
  r = R\,,\quad \theta = \Theta+\tau\,X_3\,,\quad x_3 = X_3 \,,
  \label{eq:torsion}
\end{equation}
where $(R,\Theta,X_3)$ are reference coordinates of $\mathbf{X}$, $(r,\theta,x_3)$ deformed coordinates of $\mathbf{x}$, and $\tau$ is the twist per unit length. The orthonormal cylindrical basis is $\{\mathbf{e}_r,\mathbf{e}_\theta,\mathbf{e}_3\}$. Assume the lateral surface of the cylinder is traction free, so that $\mathbf{T}\cdot\mathbf{e}_r=\mathbf{0}$ at $r=a$.

For this deformation~\eqref{eq:torsion}, the deformation gradient, left Cauchy-Green tensor and its squared tensor are
\begin{eqnarray}
  \nonumber
  \mathbf{F} &=& \mathbf{I} + \tau r\, \mathbf{e}_\theta\otimes\mathbf{e}_3\,,
  \\[2mm]
  \bB &=& \mathbf{I} +\tau^2r^2\mathbf{e}_\theta\otimes\mathbf{e}_\theta
          +\tau r(\mathbf{e}_\theta\otimes\mathbf{e}_3+\mathbf{e}_3\otimes\mathbf{e}_\theta)\,,
  \\[2mm]
  \nonumber
  \bB^2 &=& \mathbf{I}+(\tau^4r^4+3\tau^2r^2)\mathbf{e}_\theta\otimes\mathbf{e}_\theta
            +\tau^2r^2\mathbf{e}_3\otimes\mathbf{e}_3
            +(\tau^3r^3+2\tau r)(\mathbf{e}_\theta\otimes\mathbf{e}_3
            +\mathbf{e}_3\otimes\mathbf{e}_\theta)\,.
\end{eqnarray}
The invariants for pure torsion are analogous to simple shear: $i_1^\bEB = -2+\mu$, $i_2^\bEB = 1-\mu$, and $i_1^\bV = i_2^\bV = 1+\mu$, where $\mu(r) = \sqrt{4+\tau^2r^2}$. The explicit form of $\bV$ is then the same as~\eqref{eq:vexp}, swapping $\mu$ for $\eta$.

For an incompressible isotropic material, the components of the Cauchy stress tensor~\eqref{eq:cauchy} in terms of invariants of $\bEBl$ are
\begin{equation}
  \begin{aligned}
    &T_{rr} = -p + \frac{\partial\mathcal{W}}{\partial i_1^\bEB}+i_1^\bEB\frac{\partial\mathcal{W}}{\partial i_2^\bEB} \,,
  \\[2mm]
  &T_{\theta\theta} =   -p +\bigg[\frac{\partial\mathcal{W}}{\partial i_1^\bEB}+(i_1^\bEB+1)\frac{\partial\mathcal{W}}{\partial i_2^\bEB}\bigg]\bigg[1+\frac{\tau^2r^2(\mu+1)}{\mu^2+2\mu}\bigg]-\frac{\partial\mathcal{W}}{\partial i_2^\bEB}(\tau^2r^2+1) \,,
  \\[2mm]
  &T_{33} = -p + \bigg[\frac{\partial\mathcal{W}}{\partial i_1^\bEB}+(i_1^\bEB+1)\frac{\partial\mathcal{W}}{\partial i_2^\bEB}\bigg]\bigg(1-\frac{\tau^2r^2}{\mu^2+2\mu}\bigg)-\frac{\partial\mathcal{W}}{\partial i_2^\bEB}\,,
  \\[2mm]
  &T_{\theta 3} = \bigg[\frac{\partial\mathcal{W}}{\partial i_1^\bEB}+(1+i_1^\bEB)\frac{\partial\mathcal{W}}{\partial i_2^\bEB}\bigg]\frac{\tau r}{\mu}-\frac{\partial\mathcal{W}}{\partial i_2^\bEB}\tau r\,.
  \label{eq:tcomp-t}
  \end{aligned}
\end{equation}

The resultant applied moment $M$ in pure torsion can be calculated from the shear stress $T_{\theta 3}$ in~\eqref{eq:tcomp-t}, and is given by
\begin{equation}
  M = \int^{2\pi}_0\int^a_0T_{\theta 3}\,r^2drd\theta 
  = 2\pi\tau^2\int^a_0\frac{r^3}{\mu}\bigg(\frac{\partial\mathcal{W}}{\partial i_1^\bEB}
  -\frac{\partial\mathcal{W}}{\partial i_2^\bEB}\bigg)\,dr\,.
  \label{eq:moment}
\end{equation}
In order for the solid cylinder to sustain pure torsion without elongating, a resultant axial force $N$ is also required. This force can be calculated as
\begin{equation}
  \begin{aligned}
    N &= \int^{2\pi}_0\int^a_0T_{33} r\,dr\,d\theta = \pi\int^a_0(2\,T_{33}-T_{rr}-T_{\theta\theta})r\,dr
    \\[2mm]
    &= -\pi\tau^2\int^a_0r^3\bigg\{\bigg[\frac{\partial\mathcal{W}}{\partial i_1^\bEB}
    +(i_1^\bEB+1)\frac{\partial\mathcal{W}}{\partial i_2^\bEB}\bigg]\frac{i_1^\bEB+5}{\mu^2+2\mu}
    -\frac{\partial\mathcal{W}}{\partial i_2^\bEB}\bigg\}\,dr\,.
    \label{eq:axial}
  \end{aligned}
\end{equation}
For obtaining $N$ we have used the balance of linear momentum $\textrm{Div}\,\mathbf{T} = \mathbf{0}$ and traction free condition at the lateral surface, which allow to rewrite the first integral appearing in~\eqref{eq:axial} into the second one, with subsequent elimination of the pressure~\cite{truesdell2004non}. 

Compare this equation with the resultant axial force formulated in terms of invariants of $\bB$,
\begin{equation}
  N = -2\pi\tau^2\int^a_0r^3\bigg(\frac{\partial\mathcal{W}}{\partial i_1^\bB}+2\frac{\partial\mathcal{W}}{\partial i_2^\bB}\bigg)\,dr\,.
  \label{eq:axialcg}
\end{equation}
As usual, the expression~\eqref{eq:axial} written in the stretch based formulation has a more convoluted form than the Cauchy-Green based formulation~\eqref{eq:axialcg}. From the latter, Horgan and Saccomandi~\cite{horgan1999simple} pointed out that $\tau M + 2N = 0$ is a universal relation for the class of generalized incompressible neo-Hookean materials; however, no such relation between moment and axial force can be immediately inferred from~\eqref{eq:moment} and~\eqref{eq:axial} for any particular class of stretch based materials.

We now analyze how the resultant axial force of a neo-Hookean material $N_{nH}$ compares with those from a quadratic-Biot $N_q$ and a Varga $N_V$ material. By substituting their respective energy densities into~\eqref{eq:axial} and \eqref{eq:axialcg} we find
\begin{equation}
  \begin{aligned}
    N_q &= -\frac{\pi}{6\tau^2}[c_1(3\mu(a)^2+8\mu(a)-28)-9c_2](\mu(a)-2)^2\,,
    \\[2mm]
    N_V &= -\frac{\pi c_{V}}{3\tau^2}(2\mu(a)+11)(\mu(a)-2)^2\,,
    \\[2mm]
    N_{nH} &=-\frac{\pi c_{nH}}{4}\tau^2a^4\,,
  \end{aligned}
\end{equation}
where $\mu(a) = \sqrt{4+\tau^2a^2}$. This expression for $N_V$ has been previously derived in~\cite{horgan2011extension}. These can be nondimensionalized as follows: $\bar{N}_V = N_V/(\pi a^2 c_V)$, $\bar{N}_{nH} = N_{nH}/(\pi a^2c_{nH})$ and
\begin{equation}
    \bar{N}_q = \frac{N_q}{\pi a^2 c_1 } = -\frac{1}{6\tau^2a^2}[3\mu(a)^2+8\mu(a)-28-18\tilde{\gamma}\,](\mu(a)-2)^2\,,
\end{equation}
where $0 \geq \tilde{\gamma} = c_2/2c_1 > -1$. All the resultant axial forces are compressive, so that even for the quadratic-Biot material only the classic Poynting effect is observed in pure torsion, in contrast with the dual behavior found in simple shear~\eqref{eq:poy2}. We remark that the reverse Poynting effect in pure torsion can be captured by the generalized neo-Hookean model~\cite{anssari2022extension}. For the quadratic-Biot material, a response displaying transition in the Poynting effect would require $\tilde\gamma > 0$, i.e. $c_2 > 0$ (this unusual behavior in torsion has only been reported so far in pantographic metamaterials~\cite{misra2018pantographic}). From the present axial force equations, we also note that while $\bar{N}_{nH}$ presents a linear $-\tau^2a^2$ relation, the force $\bar{N}_q$ is much richer in behavior, despite both materials being quadratic in stretch.

\begin{figure}[ht]
	\centering
        \includegraphics[width=0.45\textwidth]{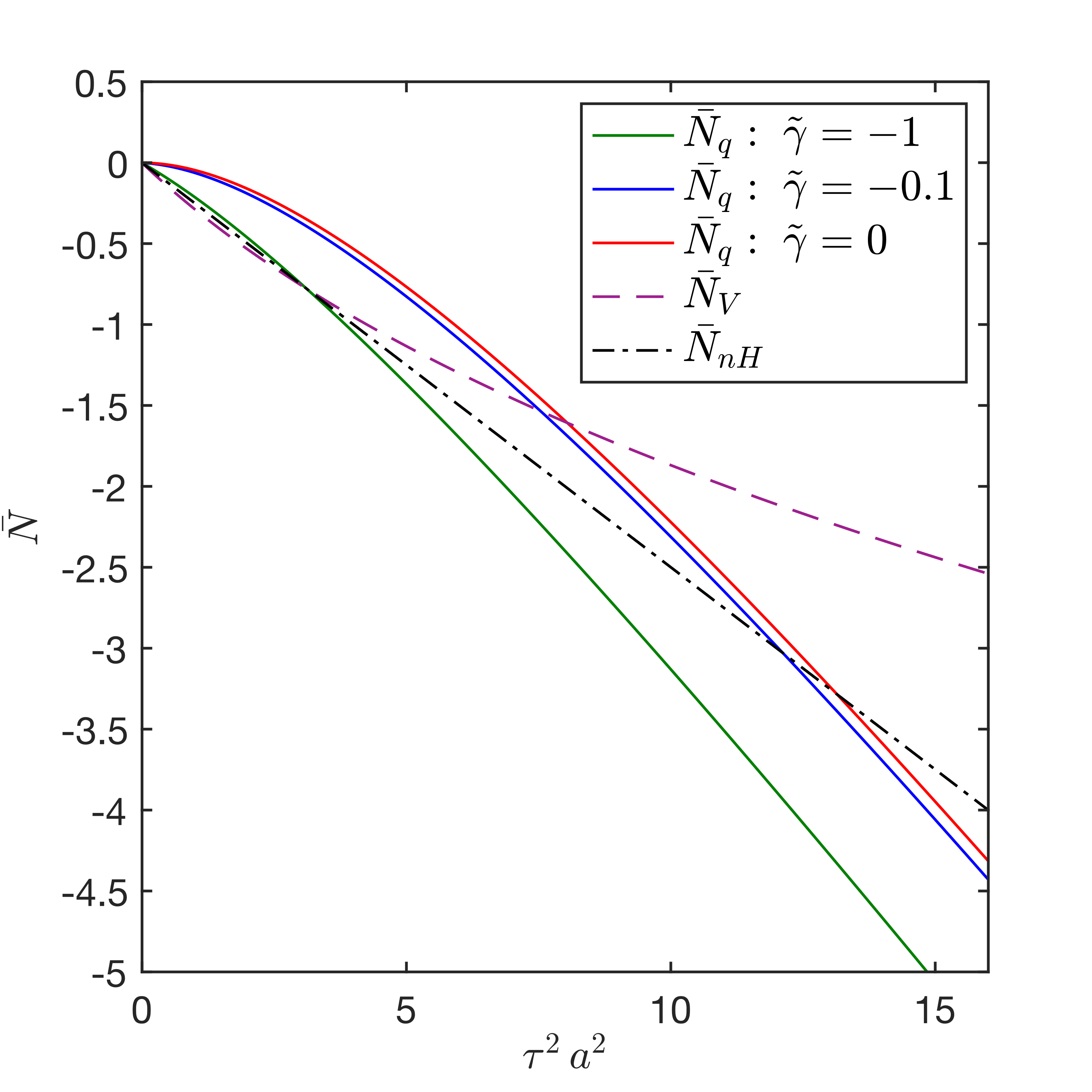}
        \includegraphics[width=0.45\textwidth]{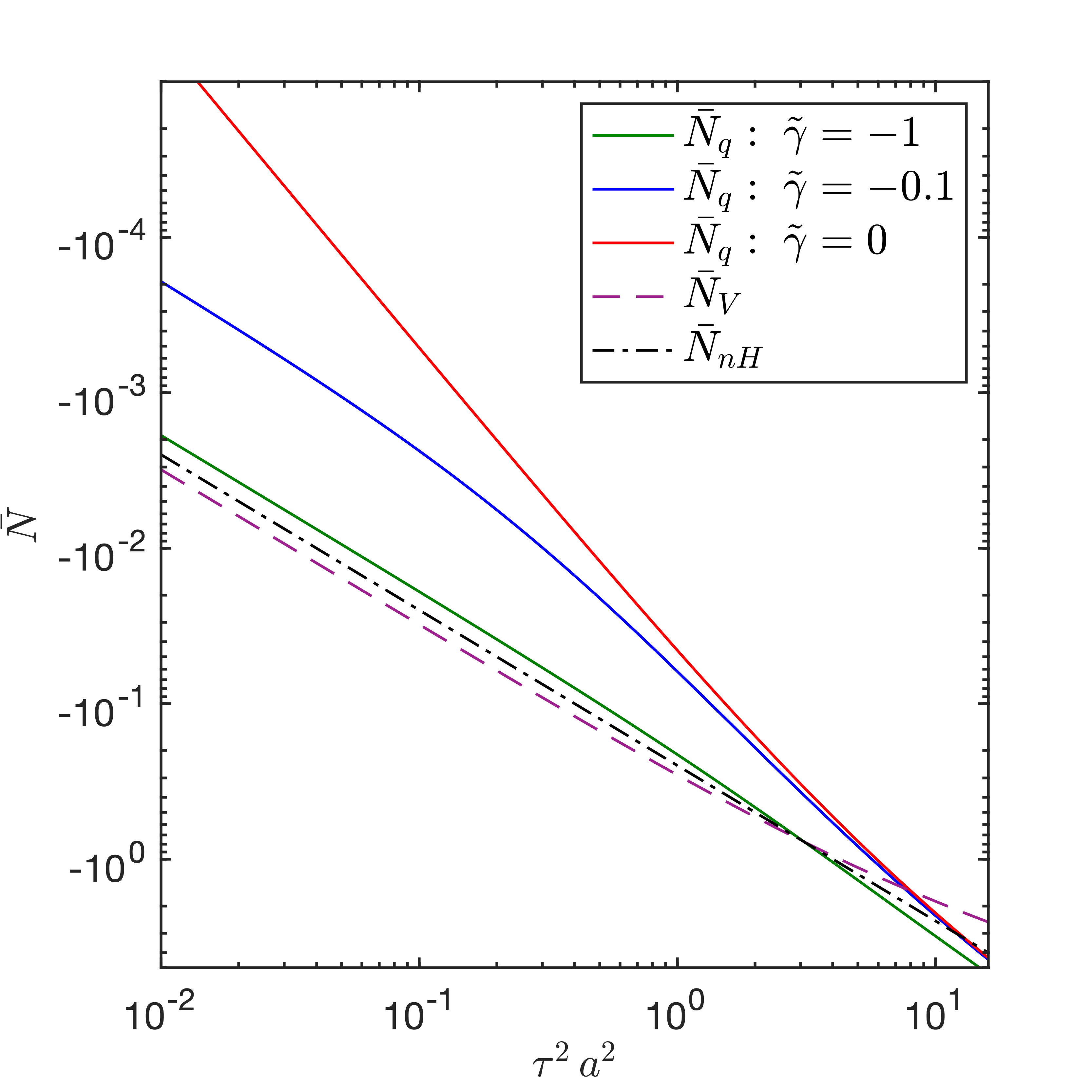}
    \caption{Nondimensional resultant axial force $\bar{N}$ as a function of the total angle of twist squared $\tau^2 a^2$. Results $\bar{N}_q$ from the quadratic-Biot model (solid line), for three different $\tilde{\gamma} = c_2/2c_1$, are compared with $\bar{N}_V$ for the Varga model (dashed line) and $\bar{N}_{nH}$ for the neo-Hookean model (dash-dotted line). The right plot presents the same data in log scale.}
	\label{fig:torsion}
\end{figure}

The nondimensional resultant axial force $\bar{N}$ is plotted in Fig.~\ref{fig:torsion} as a function of the total angle of twist squared $\tau^2 a^2$ for the quadratic-Biot with $\tilde{\gamma} = (0,\, -0.1,\, -1)$, neo-Hookean and Varga materials. Observe that for $\tau^2 a^2 < 1$, the axial force $\bar{N}$ for $\tilde{\gamma} = -1$ quadratic-Biot, neo-Hookean and Varga models present a similar $-\tau^2a^2$ behavior. For $\tau^2 a^2 > 1$ and for all allowable $\tilde{\gamma}$, the axial force $\bar{N}_q$ is proportional to $-\tau^4a^4$, so that the compressive force to maintain pure torsion increases faster than the one for a neo-Hookean material, akin results for the Gent, Fung and Horgan-Saccomandi limiting chain extensibility models shown in~\cite{kanner2008extension}. However, as $\tilde{\gamma}$ approaches zero, the quadratic-Biot material behavior becomes proportional to $-\tau^4a^4$ even for $\tau^2 a^2 < 1$, so that, for a small angle of twist, pure torsion can be supported with a much lower $\bar{N}$ than the one required for neo-Hookean or Varga materials.

\section{Conclusion}

We have contrasted classic simple functional forms (linear or quadratic in invariants) for incompressible isotropic materials on the basis of the Cauchy-Green deformation tensor with those based on the stretch $\bV $ and Bell strain $\bEBl$, which offers a different perspective on issues present in generalized neo-Hookean materials and on the importance of the second invariant of $\bB$ in nonlinear elasticity. For an isochoric simple shear deformation, any simple polynomial strain energy function of invariants of the Bell strain is shown to display the Poynting effect and shear hardening, and, for the case of a quadratic-Biot material, the second invariant of $\bEBl$ has the role of switching the Poynting effect from classic to reverse. This transition in the second normal stress may occur as a function of the amount of shear, displaying a local maximum, which is not shown by the two constant Mooney-Rivlin model. The second invariant of $\bEBl$ is also shown to be important in modeling the intensity of shear hardening displayed in simple shear.

Interestingly, using an explicit representation of $\bV$, we observe that the Cauchy stress derived from a generalized neo-Hookean material is insufficient to represent Bell strains, a primitive measure linear in stretch. This occurs because the Cauchy stress in the Cauchy-Green formulation will only present the $\bB^2$ tensor term when the energy depends on the second invariant of $\bB$, whereas an explicit expression for $\bEBl$ as a function of $\bB$ requires both $\bB$ and $\bB^2$. For pure torsion of a solid cylinder, the neo-Hookean, quadratic-Biot, and Varga materials present only the classic Poynting effect. A richer mechanical behavior is displayed by the stretch based materials in pure torsion: while the resultant axial force to support the deformation for neo-Hookean is quadratic in the total angle of twist, for the quadratic-Biot model it can change from quartic to quadratic as a function of this angle, displaying much smaller values for small twists, whereas for the Varga model the response changes from quadratic to linear.

\newpage

\section*{Acknowledgments}

The author is thankful to James Hanna for his support through the U.S. National Science Foundation grant CMMI-2001262, in addition to many helpful discussions and encouragement. The author gratefully acknowledges detailed feedback from Cornelius Horgan that helped improve the original manuscript, particularly on the pertinence of contrasting the quadratic-Biot material with Mooney-Rivlin, and is also thankful to Giuseppe Saccomandi and Jeremiah Murphy for their valuable comments. The comments of the diligent anonymous reviewers have also significantly contributed to improving the original manuscript.

\bibliographystyle{unsrturl}
\bibliography{refs_biot}

\begin{thebibliography}{10}

\bibitem{truesdell2004non}
C.~Truesdell and W.~Noll.
\newblock {\em The non-linear field theories of mechanics}.
\newblock Springer, 2004.

\bibitem{puglisi2016multi}
G.~Puglisi and G.~Saccomandi.
\newblock Multi-scale modelling of rubber-like materials and soft tissues: an
  appraisal.
\newblock {\em Proceedings of the Royal Society A: Mathematical, Physical and
  Engineering Sciences}, 472(2187):20160060, 2016.

\bibitem{destrade2017methodical}
M.~Destrade, G.~Saccomandi, and I.~Sgura.
\newblock Methodical fitting for mathematical models of rubber-like materials.
\newblock {\em Proceedings of the Royal Society A: Mathematical, Physical and
  Engineering Sciences}, 473(2198):20160811, 2017.

\bibitem{anssari2021generalised}
A.~Anssari-Benam and A.~Bucchi.
\newblock A generalised neo-hookean strain energy function for application to
  the finite deformation of elastomers.
\newblock {\em International Journal of Non-Linear Mechanics}, 128:103626,
  2021.

\bibitem{wineman2005some}
A.~Wineman.
\newblock Some results for generalized neo-{H}ookean elastic materials.
\newblock {\em International Journal of Non-Linear Mechanics},
  40(2-3):271--279, 2005.

\bibitem{horgan2012importance}
C.~O. Horgan and M.~G. Smayda.
\newblock The importance of the second strain invariant in the constitutive
  modeling of elastomers and soft biomaterials.
\newblock {\em Mechanics of Materials}, 51:43--52, 2012.

\bibitem{anssari2021central}
A.~Anssari-Benam, A.~Bucchi, and G.~Saccomandi.
\newblock On the central role of the invariant {I}2 in nonlinear elasticity.
\newblock {\em International Journal of Engineering Science}, 163:103486, 2021.

\bibitem{beatty1987class}
M.~F. Beatty.
\newblock A class of universal relations in isotropic elasticity theory.
\newblock {\em Journal of Elasticity}, 17(2):113--121, 1987.

\bibitem{horgan1999simple}
C.~O. Horgan and G.~Saccomandi.
\newblock Simple torsion of isotropic, hyperelastic, incompressible materials
  with limiting chain extensibility.
\newblock {\em Journal of Elasticity}, 56(2):159--170, 1999.

\bibitem{murphy2022inverted}
J.~G. Murphy, G.~Saccomandi, and E.~Vitral.
\newblock An inverted {R}ivlin-type universal relation for simple shear.
\newblock {\em International Journal of Non-Linear Mechanics}, 140:103911,
  2022.

\bibitem{destrade2015extreme}
M.~Destrade, M.~D. Gilchrist, J.~G. Murphy, B.~Rashid, and G.~Saccomandi.
\newblock Extreme softness of brain matter in simple shear.
\newblock {\em International Journal of Non-Linear Mechanics}, 75:54--58, 2015.

\bibitem{ogden1972large}
R.~W. Ogden.
\newblock Large deformation isotropic elasticity--on the correlation of theory
  and experiment for incompressible rubberlike solids.
\newblock {\em Proceedings of the Royal Society of London. A. Mathematical and
  Physical Sciences}, 326(1567):565--584, 1972.

\bibitem{rivlin2004note}
R.~S. Rivlin.
\newblock A note on the constitutive equation for an isotropic elastic
  material.
\newblock {\em Mathematics and Mechanics of Solids}, 9(2):121--129, 2004.

\bibitem{vitral2022quadratic}
E.~Vitral and J.~A. Hanna.
\newblock Quadratic-stretch elasticity.
\newblock {\em Mathematics and Mechanics of Solids}, 27(3):462--473, 2022.

\bibitem{hoger1984determination}
A.~Hoger and D.~E. Carlson.
\newblock Determination of the stretch and rotation in the polar decomposition
  of the deformation gradient.
\newblock {\em Quarterly of Applied Mathematics}, 42(1):113--117, 1984.

\bibitem{ting1985determination}
T.~C.~T. Ting.
\newblock Determination of {C}$^{1/2}$, {C}$^{-1/2}$ and more general isotropic
  tensor functions of {C}.
\newblock {\em Journal of Elasticity}, 15(3):319--323, 1985.

\bibitem{lurie1968theory}
A.~I. Lur\textsf{'}e.
\newblock Theory of elasticity for a semilinear material.
\newblock {\em Journal of Applied Mathematics and Mechanics}, 32(6):1068--1085,
  1968.

\bibitem{vitral2022dilation}
E.~Vitral and J.~A. Hanna.
\newblock Dilation-invariant bending of elastic plates, and broken symmetry in
  shells.
\newblock {\em Journal of Elasticity}, 2022.
\newblock \href {https://doi.org/https://doi.org/10.1007/s10659-022-09894-4}
  {\path{doi:https://doi.org/10.1007/s10659-022-09894-4}}.

\bibitem{vitral2022energies}
E.~Vitral and J.~A. Hanna.
\newblock Energies for elastic plates and shells from quadratic-stretch
  elasticity.
\newblock {\em Journal of Elasticity}, 2022.
\newblock \href {https://doi.org/https://doi.org/10.1007/s10659-022-09895-3}
  {\path{doi:https://doi.org/10.1007/s10659-022-09895-3}}.

\bibitem{irschik2009continuum}
H.~Irschik and J.~Gerstmayr.
\newblock A continuum mechanics based derivation of {R}eissner's
  large-displacement finite-strain beam theory: the case of plane deformations
  of originally straight {B}ernoulli-{E}uler beams.
\newblock {\em Acta Mechanica}, 206:1--21, 2009.

\bibitem{oshri2017strain}
O.~Oshri and H.~Diamant.
\newblock Strain tensor selection and the elastic theory of incompatible thin
  sheets.
\newblock {\em Physical Review E}, 95(5):053003, 2017.

\bibitem{wood2019contrasting}
H.~G. Wood and J.~A. Hanna.
\newblock Contrasting bending energies from bulk elastic theories.
\newblock {\em Soft Matter}, 15:2411--2417, 2019.

\bibitem{bufler1995drilling}
H.~Bufler.
\newblock On drilling degrees of freedom in nonlinear elasticity and a
  hyperelastic material description in terms of the stretch tensor. {P}art 1:
  {T}heory.
\newblock {\em Acta Mechanica}, 113(1):21--35, 1995.

\bibitem{hoger1999second}
A.~Hoger.
\newblock A second order constitutive theory for hyperelastic materials.
\newblock {\em International Journal of Solids and Structures}, 36(6):847--868,
  1999.

\bibitem{steigmann2002invariants}
D.~J. Steigmann.
\newblock Invariants of the stretch tensors and their application to finite
  elasticity theory.
\newblock {\em Mathematics and Mechanics of Solids}, 7(4):393--404, 2002.

\bibitem{beatty1992deformations}
M.~F. Beatty and M.~A. Hayes.
\newblock Deformations of an elastic, internally constrained material. {P}art
  1: {H}omogeneous deformations.
\newblock {\em Journal of Elasticity}, 29(1):1--84, 1992.

\bibitem{john1960plane}
F.~John.
\newblock Plane strain problems for a perfectly elastic material of harmonic
  type.
\newblock {\em Communications on Pure and Applied Mathematics}, 13(2):239--296,
  1960.

\bibitem{steigmann1988stability}
D.~J. Steigmann and A.~C. Pipkin.
\newblock Stability of harmonic materials in plane strain.
\newblock {\em Quarterly of applied mathematics}, 46(3):559--568, 1988.

\bibitem{mooney1940theory}
M.~Mooney.
\newblock A theory of large elastic deformation.
\newblock {\em Journal of Applied Physics}, 11(9):582--592, 1940.

\bibitem{mangan2016strain}
R.~Mangan, M.~Destrade, and G.~Saccomandi.
\newblock Strain energy function for isotropic non-linear elastic
  incompressible solids with linear finite strain response in shear and
  torsion.
\newblock {\em Extreme Mechanics Letters}, 9:204--206, 2016.

\bibitem{horgan2010simple}
C.~O. Horgan and J.~G. Murphy.
\newblock Simple shearing of incompressible and slightly compressible isotropic
  nonlinearly elastic materials.
\newblock {\em Journal of Elasticity}, 98(2):205--221, 2010.

\bibitem{destrade2012simple}
M.~Destrade, J.~G. Murphy, and G.~Saccomandi.
\newblock Simple shear is not so simple.
\newblock {\em International Journal of Non-Linear Mechanics}, 47(2):210--214,
  2012.

\bibitem{rivlin1948large}
R.~S. Rivlin.
\newblock Large elastic deformations of isotropic materials {IV}. {F}urther
  developments of the general theory.
\newblock {\em Philosophical Transactions of the Royal Society of London.
  Series A, Mathematical and Physical Sciences}, 241(835):379--397, 1948.

\bibitem{beatty1987topics}
M.~F. Beatty.
\newblock Topics in finite elasticity: {H}yperelasticity of rubber, elastomers,
  and biological tissues—with examples.
\newblock {\em Applied Mechanics Reviews}, 40(12):1699, 1987.

\bibitem{baker1954inequalities}
M.~Baker and J.~L. Ericksen.
\newblock Inequalities restricting the form of the stress-deformation relations
  for isotropic elastic solids and {R}einer-{R}ivlin fluids.
\newblock {\em Journal of the Washington Academy of Sciences}, 44(2):33--35,
  1954.

\bibitem{truesdell1952mechanical}
C.~Truesdell.
\newblock The mechanical foundations of elasticity and fluid dynamics.
\newblock {\em Journal of Rational Mechanics and Analysis}, 1:125--300, 1952.

\bibitem{mihai2011positive}
L.~A. Mihai and A.~Goriely.
\newblock Positive or negative {P}oynting effect? {T}he role of adscititious
  inequalities in hyperelastic materials.
\newblock {\em Proceedings of the Royal Society A: Mathematical, Physical and
  Engineering Sciences}, 467(2136):3633--3646, 2011.

\bibitem{liu2012note}
I.-S. Liu.
\newblock A note on the {M}ooney--{R}ivlin material model.
\newblock {\em Continuum Mechanics and Thermodynamics}, 24(4):583--590, 2012.

\bibitem{thiel2019we}
C.~Thiel, J.~Voss, R.~J. Martin, and P.~Neff.
\newblock Do we need truesdell’s empirical inequalities? on the coaxiality of
  stress and stretch.
\newblock {\em International Journal of Non-Linear Mechanics}, 112:106--116,
  2019.

\bibitem{saravanan2011adequacy}
U.~Saravanan.
\newblock On the adequacy of the existing restrictions on the constitutive
  relations to ensure reasonable elastic response of compressible bodies.
\newblock {\em Mechanics Research Communications}, 38(2):123--125, 2011.

\bibitem{janmey2007negative}
P.~A. Janmey, M.~E. McCormick, S.~Rammensee, J.~L. Leight, P.~C. Georges, and
  F.~C. MacKintosh.
\newblock Negative normal stress in semiflexible biopolymer gels.
\newblock {\em Nature materials}, 6(1):48--51, 2007.

\bibitem{destrade2015dominant}
M.~Destrade, C.~O. Horgan, and J.~G. Murphy.
\newblock Dominant negative {P}oynting effect in simple shearing of soft
  tissues.
\newblock {\em Journal of Engineering Mathematics}, 95(1):87--98, 2015.

\bibitem{araujo2020experimental}
F.~S. Ara{\'u}jo and L.~C.~S. Nunes.
\newblock Experimental study of the {P}oynting effect in a soft unidirectional
  fiber-reinforced material under simple shear.
\newblock {\em Soft Matter}, 16(34):7950--7957, 2020.

\bibitem{mihai2013numerical}
L.~A. Mihai and A.~Goriely.
\newblock Numerical simulation of shear and the {P}oynting effects by the
  finite element method: an application of the generalised empirical
  inequalities in non-linear elasticity.
\newblock {\em International Journal of Non-Linear Mechanics}, 49:1--14, 2013.

\bibitem{nunes2013simple}
L.~C.~S. Nunes and D.~C. Moreira.
\newblock Simple shear under large deformation: experimental and theoretical
  analyses.
\newblock {\em European Journal of Mechanics-A/Solids}, 42:315--322, 2013.

\bibitem{anssari2021modelling}
A.~Anssari-Benam and C.~O. Horgan.
\newblock On modelling simple shear for isotropic incompressible rubber-like
  materials.
\newblock {\em Journal of Elasticity}, 147:83--111, 2021.

\bibitem{anssari2022three}
A.~Anssari-Benam and C.~O. Horgan.
\newblock A three-parameter structurally motivated robust constitutive model
  for isotropic incompressible unfilled and filled rubber-like materials.
\newblock {\em European Journal of Mechanics-A/Solids}, 95:104605, 2022.

\bibitem{horgan2017poynting}
C.~O. Horgan and J.~G. Murphy.
\newblock Poynting and reverse {P}oynting effects in soft materials.
\newblock {\em Soft Matter}, 13(28):4916--4923, 2017.

\bibitem{dorfmann2004constitutive}
A.~Dorfmann and R.~W. Ogden.
\newblock A constitutive model for the {M}ullins effect with permanent set in
  particle-reinforced rubber.
\newblock {\em International Journal of Solids and Structures},
  41(7):1855--1878, 2004.

\bibitem{varga1966stress}
O.~H. Varga.
\newblock {\em Stress-strain behavior of elastic materials}.
\newblock Interscience, 1966.

\bibitem{rivlin1949large6}
R.~S. Rivlin.
\newblock Large elastic deformations of isotropic materials {VI}. {F}urther
  results in the theory of torsion, shear and flexure.
\newblock {\em Philosophical Transactions of the Royal Society of London.
  Series A, Mathematical and Physical Sciences}, 242(845):173--195, 1949.

\bibitem{horgan2011extension}
C.~O. Horgan and J.~G. Murphy.
\newblock Extension and torsion of incompressible non-linearly elastic solid
  circular cylinders.
\newblock {\em Mathematics and Mechanics of Solids}, 16(5):482--491, 2011.

\bibitem{anssari2022extension}
A.~Anssari-Benam and C.~O. Horgan.
\newblock Extension and torsion of rubber-like hollow and solid circular
  cylinders for incompressible isotropic hyperelastic materials with limiting
  chain extensibility.
\newblock {\em European Journal of Mechanics-A/Solids}, 92:104443, 2022.

\bibitem{misra2018pantographic}
A.~Misra, T.~Lekszycki, I.~Giorgio, G.~Ganzosch, W.~H. M{\"u}ller, and
  F.~Dell'Isola.
\newblock Pantographic metamaterials show atypical {P}oynting effect reversal.
\newblock {\em Mechanics Research Communications}, 89:6--10, 2018.

\bibitem{kanner2008extension}
L.~M. Kanner and C.~O. Horgan.
\newblock On extension and torsion of strain-stiffening rubber-like elastic
  circular cylinders.
\newblock {\em Journal of Elasticity}, 93(1):39--61, 2008.

\end{thebibliography}

\end{document}